\documentclass[12pt]{article}
\usepackage{amsmath}
\usepackage{subcaption}
\usepackage{graphicx}
\usepackage{graphics}
\usepackage{enumerate}
\usepackage{natbib}
\usepackage{xcolor}
\usepackage{appendix}
\usepackage{url}
\usepackage{multirow}
\usepackage{amsthm} % For proof
\usepackage{enumitem}
\usepackage{tabularx}
\usepackage{longtable}

\usepackage{tablefootnote}
\usepackage{longtable}
\usepackage{threeparttable}
\usepackage{bm}
\usepackage[ruled,vlined]{algorithm2e}
\usepackage{caption}
\usepackage{placeins}
\usepackage{lscape}
\usepackage{makecell}
\usepackage{booktabs}
\usepackage[margin=1.25in]{geometry}
\usepackage{pdfpages}

\usepackage{array}
\newcolumntype{L}[1]{>{\raggedright\arraybackslash}p{#1}}

\newcommand{\blind}{1}

\makeatletter
\newcommand{\pushright}[1]{\ifmeasuring@#1\else\omit\hfill$\displaystyle#1$\fi\ignorespaces}
\newcommand{\pushleft}[1]{\ifmeasuring@#1\else\omit$\displaystyle#1$\hfill\fi\ignorespaces}
\makeatother

\begin{document}

\def\spacingset#1{\renewcommand{\baselinestretch}%
{#1}\small\normalsize} \spacingset{1}

\if1\blind
{
  \title{\bf Survey-Based Estimation of Probe Group Sizes in the Network Scale-up Method: A Case Study from Jordan}
  \author{Ian Laga\\
    Department of Mathematical Sciences, Montana State University}
  \maketitle
} \fi

\if0\blind
{
  \bigskip
  \bigskip
  \bigskip
  \begin{center}
    {\LARGE\bf Survey-Based Estimation of Probe Group Sizes in the Network Scale-up Method: A Case Study from Jordan}
\end{center}
  \medskip
} \fi

\bigskip
\begin{abstract}
    Estimating the size of marginalized populations is a persistent challenge in survey statistics and public health, especially where stigma and legal restrictions exclude such groups from census and administrative data. Migrant domestic workers in Jordan represent one such population. We employ the Network Scale-up Method using the direct probe group method, estimating probe group sizes from survey respondents’ own membership rather than relying on external counts. Using data from a nationally representative household survey in Jordan, we combine the direct probe group method with Bayesian logistic mixed-effects models to stabilize small-area estimates at the Governorate level. We validate the method against census data, demonstrating that direct probe group estimates yield reliable inference and provide a practical alternative where known probe group sizes are unavailable. Our results highlight regional variation in social network size and connectivity to migrant domestic workers. We argue that the direct probe group method is more likely to satisfy the conditions required for unbiased estimation than relying on official record sizes. This work provides the first systematic validation of the direct probe group method in a small-area setting and offers guidance for adapting the Network Scale-up Method to surveys with limited sample sizes.
\end{abstract}

\noindent%
\vfill

\newpage

\spacingset{1.5} % Originally 1.5
\section{Introduction}
\label{sec:intro}

Estimating the size of marginalized populations is a central challenge in survey statistics and public health. Reliable size estimates are essential for allocating resources, designing interventions, and monitoring human rights conditions, yet these groups are often missing from census and administrative records because of stigma, legal restrictions, or informal employment. Migrant domestic workers (MDW) in Jordan are one such population, defined as non-Jordanian persons who engage in domestic work within an employment relationship. These tasks include cleaning, cooking, running household errands, caring for children, and driving for the employer, among other responsibilities. Despite their critical role in households across the country, they frequently face poor working conditions, limited mobility, and little formal recognition in official statistics \citep{IOM2020}. Developing robust methods to quantify the size of this population is therefore important both for policy and for advancing statistical approaches to population size estimation.

Estimating the size of marginalized populations is difficult. Direct enumeration via census and surveys face several issues. For example, surveys asking if respondents are MDW collected via phone, online, or face-to-face surveys may miss the population entirely because MDW are difficult to contact directly \citep{heckathorn1997respondent}. %Additionally, if MDW are living in Jordan irregularly, the household will likely not answer truthfully. Alternative size estimation approaches such as multiplier methods, capture–recapture, and respondent driven sampling avoid these specific issues, but are not feasible for MDWs in Jordan given the populations' limited appearance in administrative data, the lack of independent population lists, and the risks associated with record linkage for this vulnerable group \citep{unaids2010guidelines}.
The Network Scale-up Method (NSUM) provides an indirect approach to estimate the size of populations \citep{bernard1989estimating, mccormick2020network, laga2021thirty}. Rather than asking individuals about their own membership, respondents answer questions about people they know. The NSUM relies on collecting Aggregated Relational Data, which are answers to questions of the form ``How many people do you know who belong to group X?" where X includes ``probe groups'' with (typically) known sizes that are used to calibrate NSUM size estimates and ``target populations'' with unknown sizes that are the primary goal of the survey. By calibrating responses against the probe groups, researchers can infer the prevalence of a target population. NSUM has been applied in diverse contexts and allows researchers to study stigmatized populations without requiring direct identification \citep{bernard2010counting}.

In order to calibrate the estimates, probe group sizes and the target population estimates must exist at the same spatial resolution (e.g., city, country, region). Thus, in order to estimate target population sizes at the city level, probe group sizes must be available at the city level. Given that probe group sizes are often obtained through census records, this makes either the selection of probe groups overly restrictive or forces researchers to consider spatial resolutions larger than desired.

To simplify the probe group selection, \cite{feehan2022survey} proposed estimating the sizes of probe groups by asking respondents about their own membership in a selection of low-stigma groups, e.g., occupations. For the remainder of this manuscript, we refer to this approach as the ``direct probe group method,'' and to the resulting estimates as the ``direct probe group estimates.'' The authors were able to apply their direct probe group method to estimate the size of four target populations %(female sex workers, men who have sex with men, people who inject drugs, and people who use drugs)
in Hanoi, Vietnam, but they were only able to compare the direct probe group estimates to alternative sources for two groups: women who had a baby in the past 12 months and women who smoke. Thus, little is known about how well the direct probe group method will work in a broader range of probe groups, survey populations, and sample sizes.

This paper makes three contributions. First, we apply the direct probe group method to a nationally representative survey in Jordan, focusing on migrant domestic and hospitality workers. Second, we extend the method to small-area estimation by incorporating Bayesian logistic mixed-effects models for probe group membership. Third, we provide a systematic comparison of direct probe group estimates with census data, offering the first validation of the approach in such a setting. Based on these findings, we provide guidance on how to adapt the direct probe group method to an NSUM survey collected with a small sample, or in many sub-regions, where groups are often too small to obtain reliable direct estimates of the probe group sizes.

%While using larger subpopulations would make it more likely for respondents to belong to each probe group, it also increases recall error, negatively affecting NSUM estimates. Reliable estimates of migrant domestic and hospitality workers are particularly relevant in Jordan, where migration plays a central role in the labor market but is not fully reflected in official statistics.

The structure of the paper is as follows. In Section~\ref{sec:data}, we describe the survey and sampling design used in Jordan. Section~\ref{sec:methods} outlines our extensions to standard NSUM models to address specific data challenges. Results are reported in Section~\ref{sec:results}, where we both validate the approach and provide sub-national estimates of target population sizes. Finally, Section~\ref{sec:discussion} offers concluding remarks and directions for future work.

\section{Jordan Migrant Domestic Worker Survey}
\label{sec:data}

Jordan is administratively divided into 12 Governorates, which served as the primary sampling strata for the survey. Data were collected by the United Nations International Organization for Migration (IOM) during November and December of 2024 using a household survey design with face-to-face interviews. Within each Governorate, regions were subdivided into sampling blocks. Enumerators selected households by knocking on doors sequentially, beginning at the top floor of each building and proceeding downward. After a successful interview, enumerators skipped three doors before approaching the next household. Survey eligibility was restricted to individuals aged 18 and older, and interviews were conducted in Arabic.

A pilot study of 14 respondents in Amman was first conducted to evaluate comprehension and acceptability of the questionnaire. Feedback from the pilot led to minor wording adjustments, though no substantive changes were required.

During fieldwork, enumerators observed that elderly couples and younger women were disproportionately available at home during survey hours. To correct for this imbalance, quotas were introduced to approximate a stratified sample, with targets based on the distribution of Jordan’s male and female population across different age groups. If multiple potential respondents were available in a household, enumerators requested surveying those who belonged to the least represented gender–age group. To account for departures from the population distribution, sampling weights were constructed. Specifically, the weight for each respondent was calculated as the ratio of the expected number of respondents in their sex–age group (based on census distributions) to the actual number observed in that group within the survey sample.

The survey asked respondents how many people they knew in 35 groups, including ten related to MDW and seven related to migrant workers in the hospitality and tourism sector (MW-Hosp). For the purpose of this analysis, the following definition of MDW was included in the survey: ``Migrant domestic Workers refer to any non-Jordanian person, male or female, engaged in domestic work within an employment relationship (e.g.: working within a household and performing a variety of household services such as cleaning, household maintenance, cooking, laundry and ironing, household errands, care for children or elderly dependents, driving or guarding the house.'' The following definition of migrant workers in the hospitality and tourism sector was included in the survey: ``Migrant workers working in the hospitality or tourism sector refer to any non-Jordanian person, male or female, engaged in the hotels and tourism facilities in Jordan (hotels, airlines, travel agencies, touristic restaurants and buffets, resorts, bars and nightclub...)'' These target groups are enumerated in Appendix Table~\ref{tab:probe_groups}.

Respondents were also asked whether they currently held any of eleven selected occupations. While they could self-report other occupations, these responses are not used in the present analysis. Several of the probe group sizes are known at the Governorate-level from the Jordan Department of Statistics ``Jordan Statistical Yearbook 2023" \citep{jordanyearbook2023}. Covariates collected included gender (male or female), age (measured in years), and nationality (Jordanian, Syrian, Egyptian, or other). A full copy of the survey is included in the Supplementary Materials.

Our definition of ``knowing'' someone was as follows:

{\sffamily
\begin{quote}
We would like to ask you some questions about people that you know. People you know should satisfy all of the three following criteria:
\end{quote}
\begin{itemize}[leftmargin=5em]
    \item You know them by sight and name, and they also know you by sight and name.
    \item You have had some contact with them – either in person, over the phone, on WhatsApp, or on the computer – in the past 12 months.
    \item They live in the same Governorate you currently reside in.
\end{itemize}
\begin{quote}
These people could be your relatives, friends, colleagues, etc.
\end{quote}
}

This is the first NSUM survey to both use the direct probe group method at multiple spatial resolutions and to have reasonable known probe group sizes from census data. This combination allows us to evaluate the performance of the direct probe group estimator under our setting. The survey included nine probe groups with sizes available at the Governorate-level and nine with sizes available only at the national level or unknown entirely.

%For the remainder of the paper, we let ``Table'' refer to tables generated for this report, and ``SYB-Table'' refer to tables from the 2023 Jordan Statistical Yearbook \cite{}.

%For University student: (summing up individual univ stats), or SYB-Table 13.1.4 (or combine with 13.1.7)

%SYB-Table 2.2 provides total and gender-specific population totals for all of Jordan. We used SYB-Table 2.2 and 2.5 and the assumption that the gender-ratio was equal in all Governorates to calculate the number of men and woman aged 20-24 and 65 or older.

%Engineers are only those registered at the Jordanian Engineers Syndicate. Physicians are only those registered at the Physicians Syndicate. Lawyers are only senior advocates registered at the Jordanian Bar Syndicate.

%Respondents were asked about which occupation they held (respondents were allowed to report more than one and self-report ``Other'', although the ``Other'' categories were not used in this analysis) and whether they were married or went through a divorce in 2023. Respondents were also asked 13 questions about migrant domestic workers across different genders and nationalities and six questions about migrant workers in the hospitality sector. 

\section{Methodology}
\label{sec:methods}

We fit a modified form of the correlated NSUM model \citep{laga2023correlated}. The correlated NSUM model (using the authors' notation) assumes that the prevalence for group $k$ is given by $p_k = \exp(\rho_k) = N_k / N$, where $N_k$ is the size of group $k$ and $N$ is the total population size. We collect survey responses $y_{i,k}$ for respondents $i \in \{1, \ldots, n\}$ and for groups $k \in \{1, \ldots, K\}$. Given the Governorates $g \in \{1, \ldots, G\}$, we modify this slightly, such that $p_{g,k} = \exp(\rho_{g,k}) = N_{g,k} / N_g$, where $N_{g,k}$ now represents the size of group $k$ within Governorate $g$ and $N_g$ denotes the total population size of Governorate $g$. This approach treats each prevalence for a group as a random effect, allowing for pooling across Governorates and a shrinkage towards a common mean for Governorates with small sample sizes. We also let $d_i = \exp(\delta_i)$ represent a respondent's degree, or social network size, i.e., the number of people they know according to the definition above. Prior to analysis, responses were truncated at 100 due to a large number of unreasonably large values. Truncating the responses at an assumed maximum value is consistent with existing NSUM analyses \citep{zheng2006many, vogel2025accounting}.

Then, our modified correlated NSUM model is given by
\begin{alignat}{2}
y_{i,k} &\sim \text{Poisson}\left(\exp\left\{\delta_i + \rho_{g,k} + \bm{\beta} \bm{z}_{i} + b_{i,k} \right\} \right), &\quad& \label{eq:correlated} \\
\delta_i &\sim \mathcal{N}(0, \sigma_\delta^2), &\quad& \sigma_\delta \sim Cauchy_+(0, 2.5), \nonumber \\
\rho_{g,k} &\sim \mathcal{N}(\mu_{\rho,k}, \sigma^2_{\rho,k}), &\quad& \mu_{\rho,k} \sim \mathcal{N}(\mu_{\rho,\text{base}}, 100), \nonumber \\
\sigma_{\rho,k} &\sim \mathcal{N}_+(\sigma_{\rho,\text{base}}, 100), &\quad& \sigma_{\rho,\text{base}} \sim Cauchy_+(0, 2.5), \nonumber \\
\mu_{\rho,\text{base}} &\sim \mathcal{N}(0, 100), &\quad& \beta_{pk} \sim \mathcal{N}(0, 100), \nonumber \\
\bm{b}_i &\sim \mathcal{N}_{K}(\bm{\mu}, \Sigma_{K\times K}), &\quad& \tau_{N,k} \sim Cauchy_+(0, 2.5), \nonumber \\
\mu_k &= \log\left(\frac{1}{\sqrt{1 + \tau_{N,k}^2}}\right), &\quad& \tau_k = \sqrt{\log(1 + \tau_{N,k}^2)}, \nonumber \\
\Sigma &= \text{diag}(\bm{\tau}) \, \Omega \, \text{diag}(\bm{\tau}), &\quad& \Omega \sim \text{LKJ-Corr}(2) \nonumber
\end{alignat}
where $\beta_{pk}$ represents the coefficient corresponding to the covariate $p$ for group $k$ recorded in $\bm{z}_i$, and $b_{i,k}$ represents the bias associated with respondent $i$ towards group $k$. As in \cite{laga2023correlated}, the bias structure is fixed such that $E(e^{b_{i,k}}) = 0$. We assume the covariance matrix of the biases, given by $\Sigma$, is the same for all Governorates, and is scaled according to variance vectors $\bm{\tau}$, where LKJ-Corr refers to the Lewandowski-Kurowicka-Joe prior for correlation matrices. While this assumption may not hold precisely in practice, it would be difficult to estimate the covariance matrix in Governorates with small sample sizes, so we choose to assume a shared matrix.

When evaluating the posterior distribution, the probability model is weighted by the sampling weights $w_i$, i.e.,
\begin{equation*}
    p(\bm{\theta} | \bm{y}, \bm{z}, \bm{x}) \propto \left[\prod_i w_i p(\bm{y}_i | \bm{\theta}, \bm{z}, \bm{x})\right] p(\bm{\theta}),
\end{equation*}
where $p(\bm{\theta} | \bm{y}, \bm{z}, \bm{x})$ is the posterior distribution, $p(\bm{y}_i | \bm{\theta}, \bm{z}, \bm{x})$ is the probability model for the responses for respondent $i$, and $p(\bm{\theta})$ is the prior distribution for the set of parameters $\bm{\theta}$.

As for most Bayesian NSUM models, the $\delta_i$ and $\rho_{g,k}$ are non-identifiable, meaning post-hoc scaling is necessary to produce meaningful degree and prevalence estimates \citep{zheng2006many}. Informally, this scaling works by centering prevalence estimates around the known prevalences. We combine this post-hoc scaling with the direct probe group method.

The direct probe group method first requires estimating the prevalences of the probe groups from the membership questions \citep{feehan2022survey}. In our survey, several probe groups were too small given the sample size within each Governorate, leading to situations where no respondents reported membership, resulting in zero-valued size estimates. To address this limitation, we first estimate probe group prevalences by fitting independent Bayesian logistic mixed-effects models, each including a random intercept for Governorate. This hierarchical structure allows information to be shared across Governorates, yielding stable prevalence estimates even for sparsely represented groups and mitigating the problem of zero counts. Since the Governorates populations $N_{g}$ are known, we can estimate both the size, $N_{g,k}$, and the prevalence, $\gamma_{g,k} = N_{g,k} / N_{g}$, for each probe group. The logistic mixed-effects model produces posterior samples for $\gamma_{g,k}$ given by $\gamma_{g,k}^{\{1\}}, \ldots, \gamma_{g,k}^{\{R\}}$.

While \cite{feehan2022survey} proposed an incorporated unknown probe group size model which included a distribution on $Z_g$ (originally the total number of survey respondents who report being members of probe group $g$), consistent with most Bayesian NSUM models, these probe group sizes are used only in post-hoc scaling. Additionally, we are able to incorporate the uncertainty of the direct size estimates into the scaling using the posterior samples for $\gamma_{g,k}$. Specifically, the combined scaling is given by the general Algorithm~\ref{alg:bootstrap_scale}, which produces scaled log-prevalence samples $\tilde{\rho}_{g,k}^{\{m,r\}}$ and scaled log-degree samples $\tilde{\delta}_{i}^{\{m,r\}}$ which incorporate the uncertainty of the direct probe group estimates. Note that the function for $c^{\{m,r\}}_{g,k}$ can be replaced by any scaling function, like those proposed by \cite{zheng2006many} and \cite{laga2023correlated}. In this manuscript, we use the scaling procedure proposed in \cite{laga2023correlated} where each population is weighted equally, rather than by their population size.

\begin{algorithm}[H]
\SetAlgoLined
\KwResult{Scaled log-prevalence and log-degree estimates}
    Set $M$ and $R$ equal to the number NSUM and logistic regression posterior samples, respectively\\
    Set $\mathcal{K}$ equal to the set of probe groups, where $|\mathcal{K}|$ is the number of probe groups\\
    For each Governorate $g$:
    \begin{enumerate}
        \item For $m \in \{1, \ldots, M\}$, take the $m^{th}$ draw of $\rho_{g,k}$, denoted $\rho_{g,k}^{\{m\}}$, $\forall\, k$:
        \begin{enumerate}
            \item For $r \in \{1, \ldots, R\}$, take the $r^{th}$ draw of $r_{g,k}$ for all probe groups with direct probe group estimates from the logistic regression model, denoted $\gamma_{1,k}^{\{r\}}$ through $\gamma_{G,k}^{\{r\}}$. For groups without direct probe group estimates, let $\gamma_{g,k}^{\{r\}}$ equal the known sizes:
            \begin{enumerate}
                \item Compute the shift constant:
                \begin{equation*}
                    c^{\{m,r\}}_{g} = \log \left[ \frac{1}{|\mathcal{K}|} \sum_{k \in \mathcal{K}} \left( \frac{exp\{\rho_{g,k}^{*(b)}\}}{\gamma_{g,k}^{\{r\}}}\right) \right]
                \end{equation*}
                \item Apply the scaling adjustment:
                \begin{align*}
                \tilde{\rho}_{g,k}^{\{m,r\}} &= \rho_{g,k}^{\{m\}} - c^{\{m,r\}}_{g} \quad{\forall\, k}\\
                \tilde{\delta}_{i}^{\{m,r\}} &= \delta_{i}^{\{m\}} + c^{\{m,r\}}_{g} \quad{\forall\, i \in g}
                \end{align*}
            \end{enumerate}
        \end{enumerate}
    \end{enumerate}
    Return $\tilde{\rho}_{g,k}^{\{m,r\}}$ and $\tilde{\delta}_{i}^{\{m,r\}}$ as the scaled posterior samples of the log-prevalence and log-degree, respectively
\caption{Direct probe group size bootstrap scaling procedure}
\label{alg:bootstrap_scale}
\end{algorithm}

Note that while $\rho_{g}^{\{m\}}$, for example, is of length $M$, the resulting scaled values $\tilde{\rho}_{g}^{\{m,r\}}$ are of length $M * R$ as the uncertainty from the logistic regression model is distributed across all NSUM posterior samples. Then, estimates and uncertainty intervals for the prevalences are given by summarizing $\exp\left(\tilde{\rho}_{g}^{\{m,r\}} \right)$. For known probe group sizes, one may set $N_{g,k}^{\{r\}}$ equal to the known size, although as we discuss later in Section \ref{sec:dir_est}, there may be advantages to using the direct probe group estimates in place of the known sizes. For groups like ``Died in 2023'', it is impossible to estimate the direct probe group size, so the known size must be used directly.

We found that the extra uncertainty from the direct probe groups combined with the Governorate-level analysis makes summarizing $\exp\left(\tilde{\delta}_{i}^{\{m,r\}} \right)$ by the posterior mean problematic. Specifically, exponentiating the wider intervals produces unreasonable estimates. We offer two solutions. The first approach scales the log-degrees by averaging the estimated probe group prevalences over all $R$ draws prior to computing the shift constant. This approach removes the uncertainty in the probe group estimates, leading to substantially smaller posterior densities. The second approach uses posterior medians to summarize these degree estimates. This stabilizes the point estimates but leads to wider than expected credible intervals. While the credible intervals reflect the uncertainty in our model, they are often wider than is usable in practice (e.g., one Governorate produced a 95\% credible interval for the average degree from 55 to 1605, with a point estimate of 430). Based on these findings, we believe it is more practical to scale the log-degrees after averaging the direct probe group estimates and produce anti-conservative credible intervals. We present degree estimates in this manuscript using these anti-conservative intervals.

%Unlike existing scaling methods, Algorithm 1 does not scale $\delta_i$ using a common shift constant. We found that scaling $\delta_i$ separately for each Governorate produced unreliable and extreme estimates. Instead, to obtain Governorate-level degree estimates, we scale $\delta_i$ by summing the estimated prevalences and probe group sizes over all Governorates. Although these degree estimates are no longer tied exactly to the Governorate-level population sizes, they provide a reasonable proxy.

%and log-degree samples $\tilde{\delta}_i^{\{m,r\}}$

%and $\exp\left(\tilde{\delta}_i^{\{m,r\}} \right)$

%$\tilde{\delta}_i^{\{m,r\}} = \delta_i^{\{m\}} + c^{\{m,r\}}_{g} \quad{\text{for respondents in Governorate $g$}}$

\subsection{Probe Group Definitions}

We show that probe group sizes do not need to match the “true” sizes exactly for the model to perform well and that aligning probe groups with respondents’ understanding of the groups can improve accuracy. Traditional definitions of groups, such as those used in census reports, often differ from how respondents interpret group membership. For example, the Jordan Statistical Yearbook reports the number of physicians employed at the Ministry of Health, a category that includes veterinarians, dentists, residents, general practitioners, and specialists. However, survey respondents are unlikely to interpret \textit{physicians} in this way. If the question mentions only \textit{physicians}, they may exclude veterinarians; if it lists every subgroup and specifies ``employed at the Ministry of Health,'' respondents may be uncertain who qualifies or even what a \textit{resident} is. In either case, their answers systematically deviate from the official counts. Using these official sizes for scaling would therefore bias target population estimates. While direct probe group size estimates may differ from some underlying ``true'' size, we argue that anchoring on respondents’ interpretation may yield more reliable results.

We more formally discuss the effect probe group sizes have on degree and size estimators through theory developed for design-based estimators \citep{feehan2016generalizing, feehan2022survey}. Given the complexity of the model-based and Bayesian models, we offer no proofs that the theory holds under our setting, but rather informally posit that the same benefits of the direct probe group method extends to the Bayesian models.

\cite{feehan2022survey} note three conditions to produce unbiased degree and size estimates when using the direct probe group method:
\begin{enumerate}
    \item $x_{\mathcal{A}} = \sum_{i \in F} x_i = N_{\mathcal{A}}$
    \item $y_{F,\mathcal{A}} = d_{F,\mathcal{A}}$
    \item $\bar{d}_{\mathcal{A},F} = \bar{d}_{F,F}$
\end{enumerate}
Condition 1 states that that the number of people in the frame population who report belonging to the probe groups $\mathcal{A}$, denoted $x_{\mathcal{A}}$ is equal to the size of the probe groups, $N_{\mathcal{A}}$. Condition 2 states respondents accurately report their connections to the probe groups. Condition 3 states the average degree from the probe groups to the frame population is the same as the average degree from the frame population to the frame population.

We argue that in practice, respondents report membership to $\mathcal{A}'$, a perturbed set of probe groups, and thus, $x_{\mathcal{A}'} \neq N_{\mathcal{A}}$ outside of random chance. While this seems problematic, note that respondents will likely also report connections to $\mathcal{A}'$ (while this assumption would be violated if respondents were asked about their membership in stigmatized/criminal probe groups, we assume the probe groups are chosen carefully to ensure accurate reporting). Thus, the probe group method remains unbiased by replacing \textit{all} instances of $\mathcal{A}$ with $\mathcal{A}'$ (Condition 3 is difficult to ensure in practice, but has roughly the same likelihood of holding under $\mathcal{A}'$ as it does under $\mathcal{A}$). In fact, estimators which rely on known population sizes are likely to have a mismatch between $\mathcal{A}$ and $\mathcal{A}'$ unless on average, the respondents' understanding of the probe groups happens to coincide with the definitions used to calculate the known population sizes.

While the estimators are theoretically unbiased, the necessary conditions are thus often ignored in the existing literature. %\cite{habecker2015improving} propose removing poorly performing probe groups in leave-one-out back-estimation. However, back-estimation may appear unbiased while still violating this condition.
\cite{mccarty2001comparing} noted that recall error is connected to respondent's uncertainty about ambiguous group definitions. The authors commented that this ambiguity is unavoidable since probe groups are limited to those with official counts. The direct probe group estimator solves this problem directly, with the caveat that the target population size estimate corresponds to respondents' understanding of the target population, which may differ from the researcher's original intent. This limitation can be addressed by very clearly defining the boundary of the target population.

\section{Results}
\label{sec:results}

In the analysis of the Jordan NSUM survey, the covariate effects, $\bm{\beta}$, were allowed to vary for each subpopulation, i.e., age was allowed to affect the Poisson rate differently for each group. Two respondents refused to report their exact age, so their age was estimated as the mean of all respondents who shared the same age group. The baseline rate corresponds to Jordanian females with mean age of the respondents (41.9). To produce final size estimates, we added one half of the male coefficient, averaging estimates between male and female responses.

%The total population of Jordan is $11,516,000$. To calculate the population of Jordan aged 18 and older, we assume $2/5$ of the population aged 15-19 are aged 18 or 19. The male and female populations of each Governorate are given in the 2023 Jordan Statistical Yearbook \citep{jordanyearbook2023}.

For our analysis, the total population of Jordan is $11,516,000$, where $N = 6,873,239$ represents the number of people aged 18 or older in Jordan, calculated by summing up the age groups in the Statistical Yearbook and assuming the number of 18 and 19 year-olds is equal to $2/5$ of the population of $15-19$ year-old. While the actual number of 18 and 19 year-olds will be slightly different, this approximation serves as our best estimate given the available data and should introduce only minor bias in the results. The number of males and females living in Jordan within each age-group are obtained using the same approach, but with the sex-specific values from the Statistical Yearbook. %SYB-Table 2.5.

Using these sex/age population sizes, we calculate the sample weight for each respondent as
\begin{equation*}
    w_{i} = \frac{N_{s,a} \cdot n \cdot N_g}{N \cdot O_{s,a,g}},
\end{equation*}
where $N_{s,a}$ is the number of Jordan adults of sex $s$ and age group $a$, $n$ is the observed sample size, $N_g$ is the adult population size of Governorate $g$, $N$ is the adult population size of Jordan, $O_{s,a,g}$ is the observed count in the sample corresponding to the sex $s$/age $a$/Governorate $g$ combination.

\subsection{Direct Probe Group Estimates}
\label{sec:dir_est}

\begin{figure}[t!]
\centering\includegraphics[width=\textwidth]{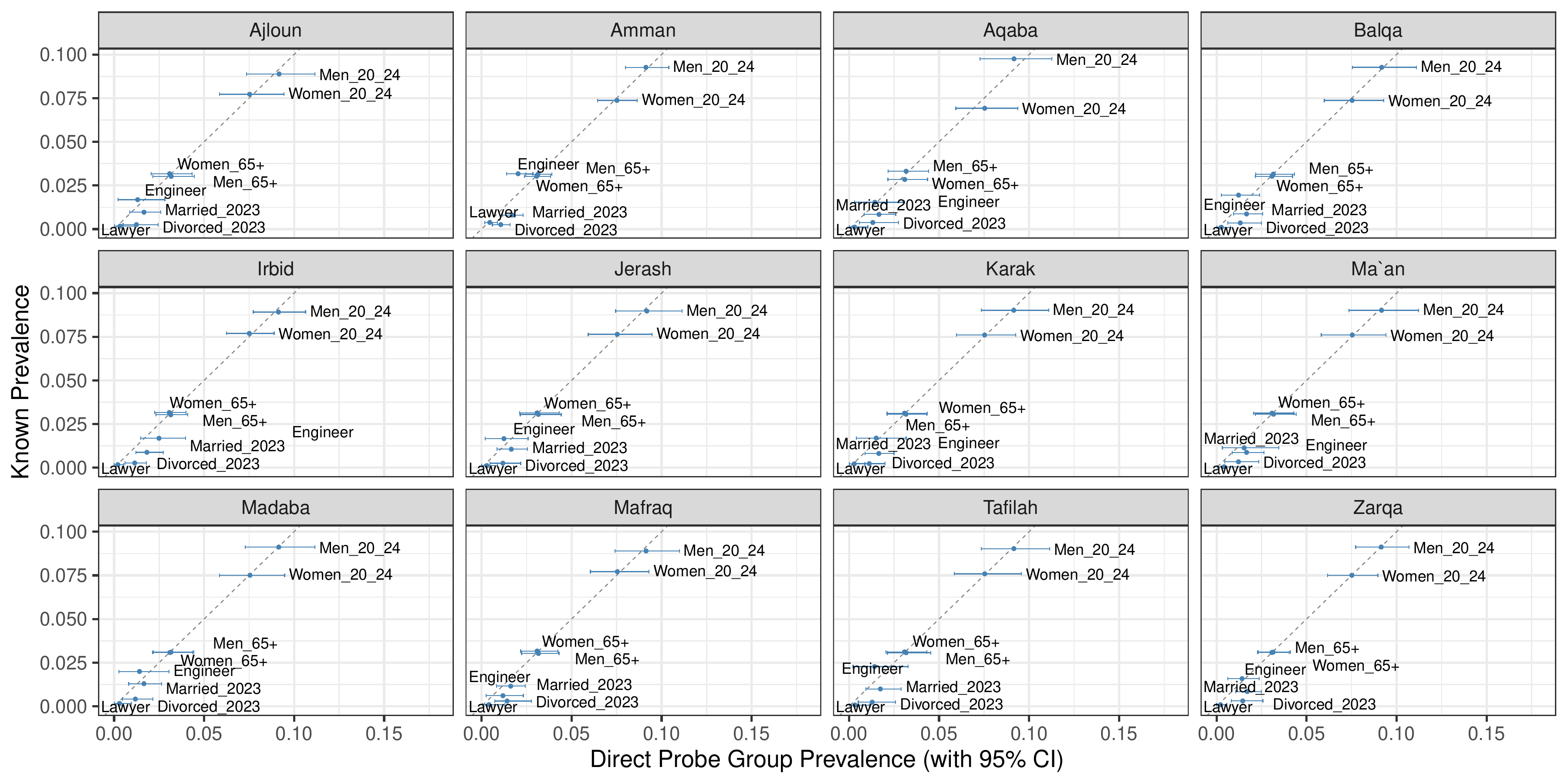}
\caption{Governorate-level probe group sizes from the Ministry of Labor against the direct probe group estimates and 95\% credible intervals.}
\label{fig:direct_vs_mol}
\end{figure}

Figure~\ref{fig:direct_vs_mol} plots the direct probe group size estimates and corresponding 95\% credible intervals against the known probe group sizes from the Statistical Yearbook for each Governorate. The performance is relatively consistent across all Governorates. We note that while the performance of the direct probe group estimator is notably worse for groups with small prevalences, it's difficult to determine whether the error is due to low prevalences (and thus high variation in respondent membership) or a known NSUM bias like recall error. While we argued above that differences between ``known'' and direct probe group sizes may not matter, we find here the estimated sizes from the mixed-effect model are able to accurately recover the true prevalences, even for populations like \textit{Lawyer} where some Governorates had zero-reported members of that occupations. Therefore, in this analysis, using the direct probe group method or the known sizes likely only has a small effect on the degree and size estimates.

\subsection{Degree Estimates}

We estimate that people living in Irbid and Karak have the largest social networks (means of 500 and 510, respectively), while those living in Ma`an and Aqaba have the smallest social networks (means of 255 and 185, respectively). Furthermore, the 95\% credible intervals do not overlap between the Governorates with the largest and smallest social networks. The estimates and credible intervals for all districts are shown in Table~\ref{tab:degrees}. These results highlight the importance of the small-area estimation approach. While respondents all live in Jordan, there are key differences between Governorates that are lost when sub-national information is ignored.

\begin{table}[!t]
\centering
\caption{Mean (2.5\% and 97.5\% quantile) of posterior mean social network sizes of Jordanians (general population) and migrant domestic workers (MDW) by Governorate. All values are rounded to the nearest 5.}
\label{tab:degrees}
\begin{tabular}{lrr}
\toprule
Governorate & General Population & MDW \\
\midrule
Ajloun & 425 (385–465) & 60 (55–65) \\
Amman & 260 (245–275) & 75 (70–80) \\
Aqaba & 185 (165–210) & 100 (95–105) \\
Balqa & 345 (315–380) & 75 (65–80) \\
Irbid & 500 (470–530) & 75 (70–80) \\
Jerash & 410 (375–445) & 80 (75–85) \\
Karak & 510 (465–555) & 65 (60–70) \\
Ma`an & 255 (210–300) & 100 (95–105) \\
Madaba & 390 (355–435) & 70 (65–80) \\
Mafraq & 355 (325–390) & 75 (70–80) \\
Tafilah & 420 (365–485) & 80 (50–105) \\
Zarqa & 310 (290–335) & 80 (75–85) \\
\bottomrule
\end{tabular}
\end{table}

\subsection{Covariate Estimates}

\begin{figure}[t!]
\centering\includegraphics[width=0.95\textwidth]{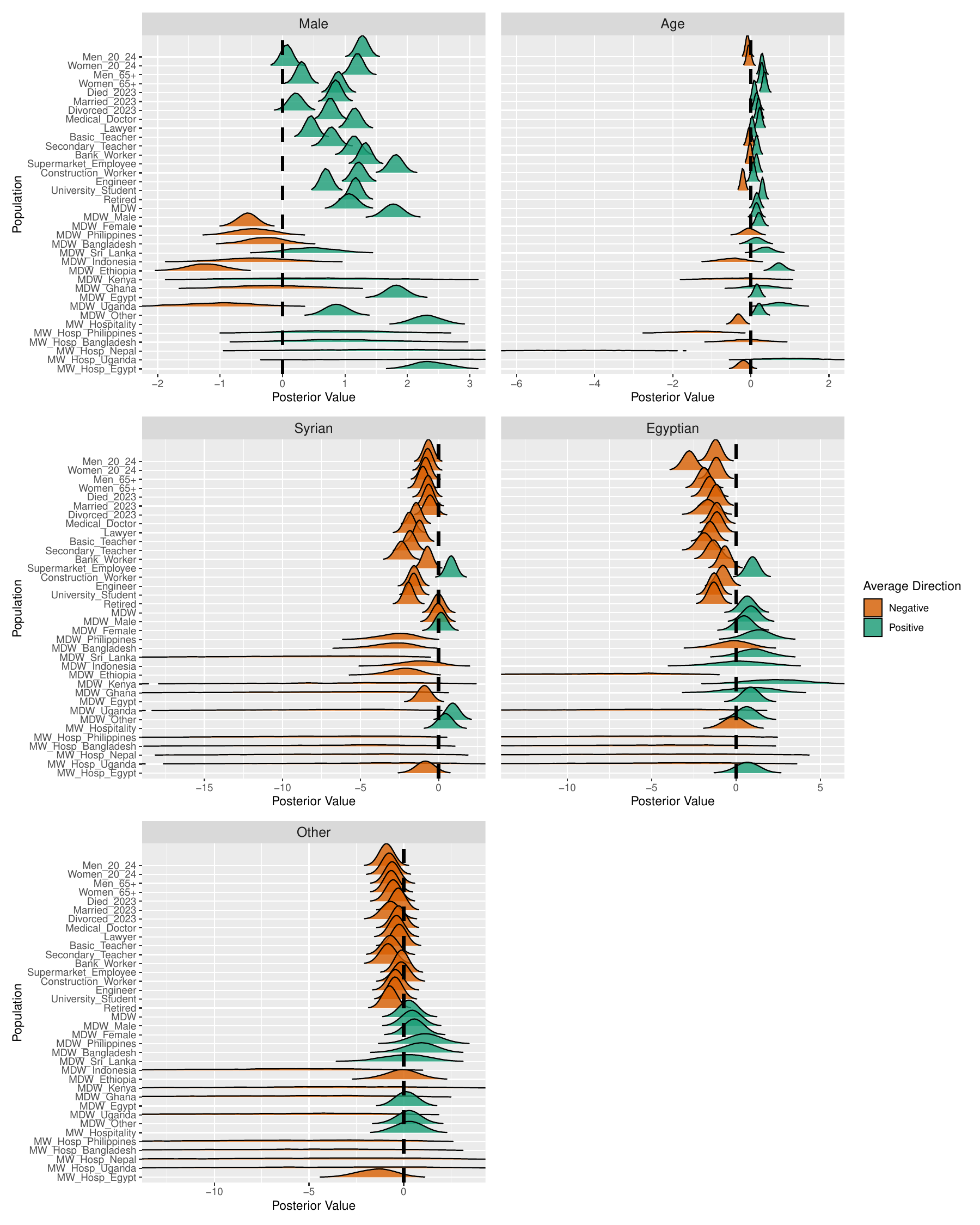}
\caption{Posterior samples of covariate effects for respondent gender, age, and nationality on each population.}
\label{fig:cov_ridge}
\end{figure}

The inclusion of covariates in the model allows us to better understand how survey respondents in Jordan form social networks. The estimated covariate-effects are shown in Figure~\ref{fig:cov_ridge}. For example, the plot suggests men have, on average, much larger social networks than women. Men also tend to be more connected to Male MDW, while women tend to be more connected to Female MDW. Men are especially connected to construction workers, while women are especially connected to women aged 20-24. We also find that older respondents are more connected to people who died in 2023 and retired individuals, while younger respondents are more connected to younger men and women and university students.

Concerning nationalities, individuals from Syria and Egypt are highly connected to construction workers and MDW, and Syrians living in Jordan seem to be especially socially disconnected from lawyers, secondary teachers, and bank workers.

\subsection{Correlation Estimates}

\begin{figure}[t!]
\centering\includegraphics[width=0.95\textwidth]{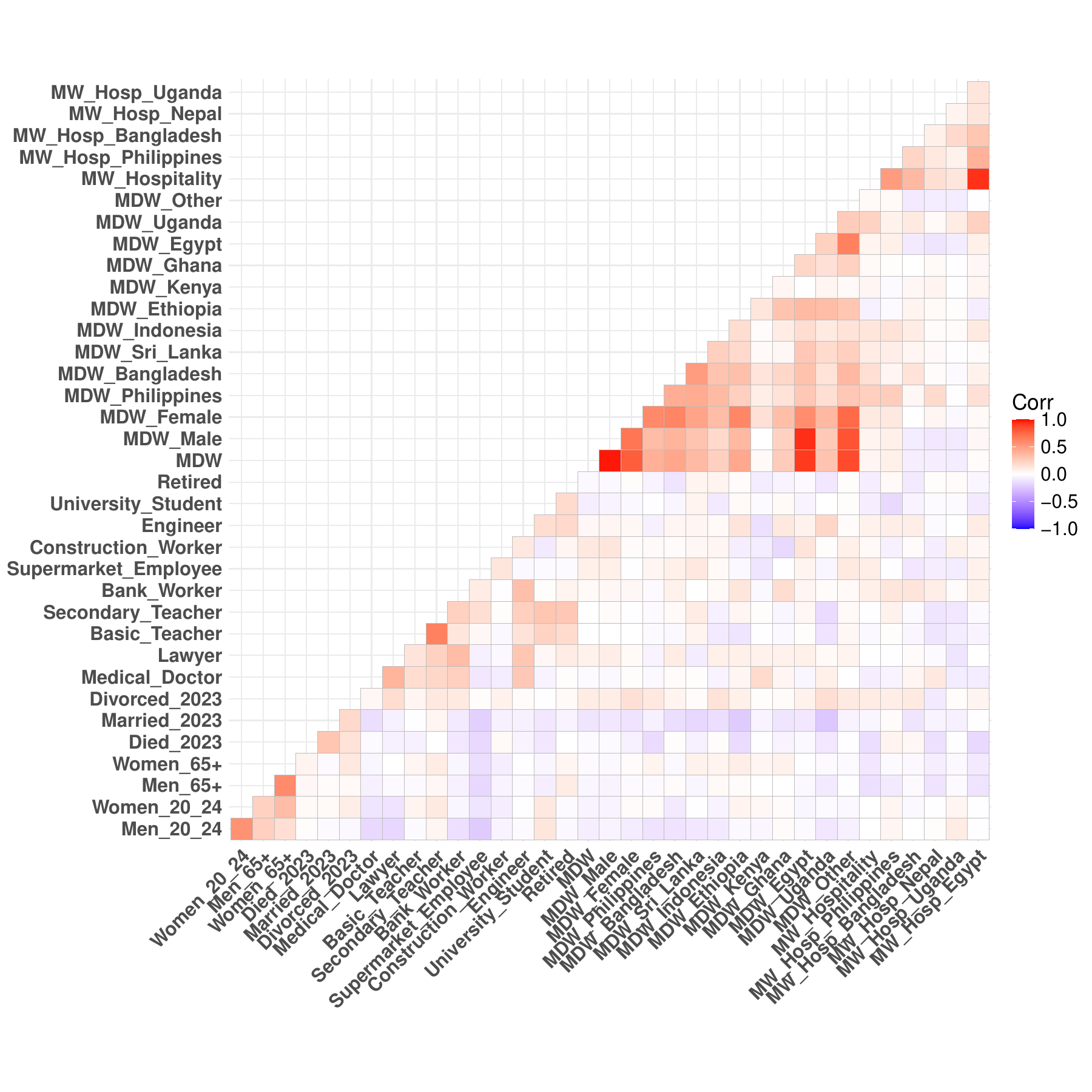}
\caption{Estimated correlation between responses for members of the General Population Survey}
\label{fig:genpop_corr}
\end{figure}

Figure~\ref{fig:genpop_corr} shows the estimated correlation between groups. The correlations typically match our expectations, including high correlation between basic and secondary teachers, men and women aged 65+, and secondary teachers and engineers and university students. Interestingly, we find an estimated negative correlation between supermarket employees and most other probe groups. Two alternative interpretations are that supermarket employees have substantially smaller social network sizes than the remaining probe groups or that supermarket employees have distinct social networks from the remaining groups.

From the model, we find respondents who know more MDW of one group tend to know more MDW of other groups. Despite not being included as a direct question on the survey, the relatively large correlations between Female MDW and MDW from Bangladesh, Sri Lanka, and Ethiopia, when compared to the correlation between Male MDW and these countries, suggests that MDW from these countries are primarily female, while MDW from Egypt are primarily male. This relationship is consistent with the Jordanian Ministry of Labor work permits.

\subsection{Migrant Domestic Workers Estimates}

\begin{figure}[t!]
\centering\includegraphics[width=\textwidth]{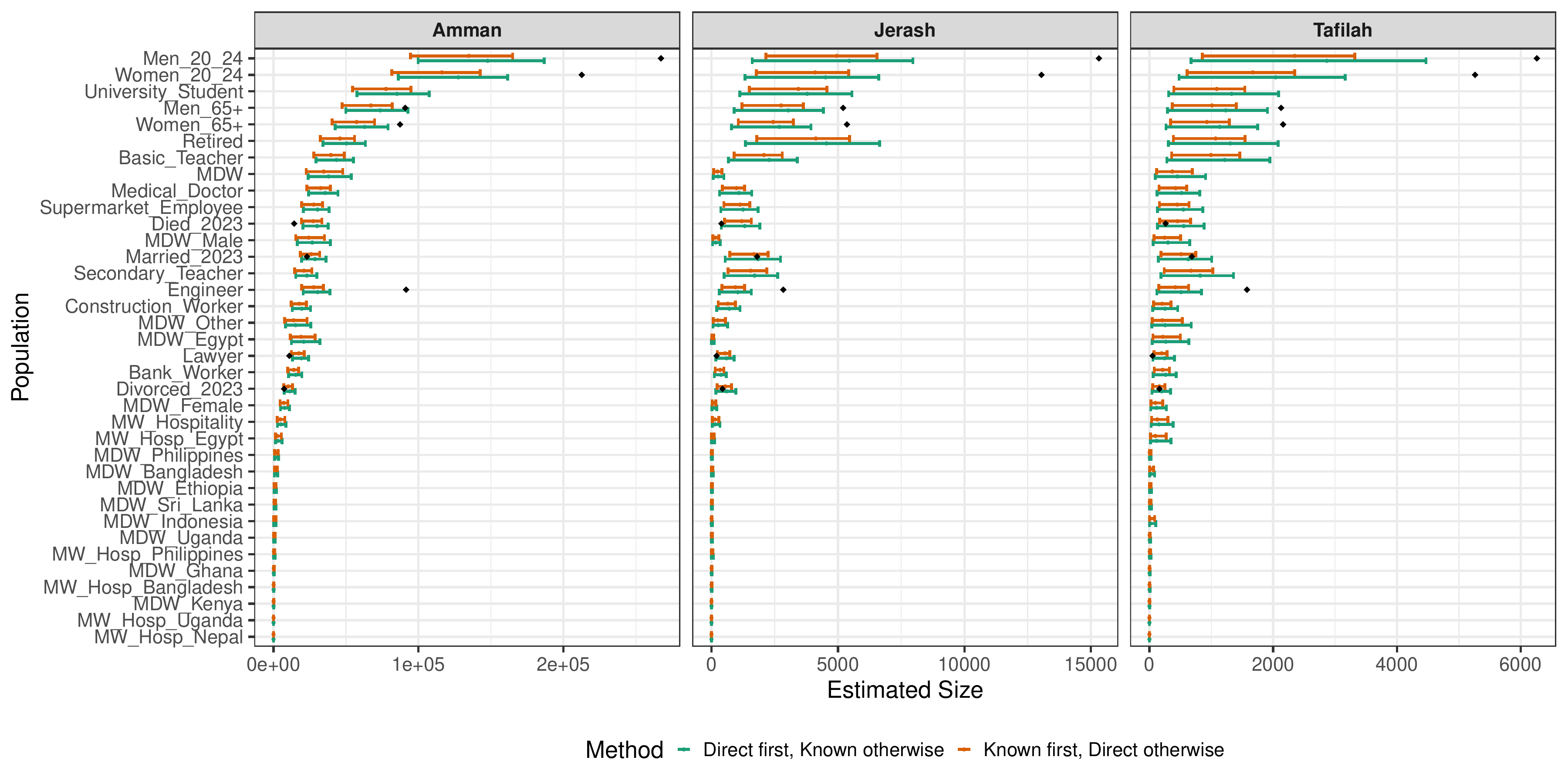}
\caption{Posterior mean and 95\% credible intervals for all populations from the larger (Amman), medium (Jerash), and small (Tafilah) Governorates. The ``direct first, Known otherwise'' method (green) denotes the method using all direct probe group estimates when available, whereas the ``Known first, Direct otherwise'' method (orange) uses all known probe group sizes when available. The known size is shown as a black diamond when available.}
\label{fig:direct_vs_mol_post}
\end{figure}

We present MDW size estimates within each Governorate both using only the direct probe groups estimates and using the known probe group sizes when available and the direct probe group estimates when not available. The estimates and 95\% credible intervals in Figure~\ref{fig:direct_vs_mol_post} shows that while the posterior density of the two methods are similar, the direct first method produces larger credible intervals, as expected, with slightly larger point estimates as well.

Although it is difficult to determine whether known sizes or direct probe group estimates yield more accurate results, these findings suggest that researchers can confidently use the direct probe group method when known sizes are unavailable. This approach is unlikely to produce dramatically different or irreconcilable estimates compared to those based on known population sizes.

\section{Discussion}
\label{sec:discussion}

We find the direct probe group estimator especially useful for estimating population sizes at smaller spatial resolutions, where reliable probe group sizes are often unavailable. However, our results suggest that careful attention to probe group prevalence is crucial. Previous recommendations favored very rare groups, with prevalences as low as 0.1–0.2\%, under the assumption that larger groups would exacerbate recall error \citep{mccormick2010many}. In practice, very small groups often lead to sparse data. For example, with 500 respondents, there is still over a 36\% chance that no one reports membership in a group with a prevalence of 0.2\%. Based on our findings, for a sample size of around 500, we recommend selecting probe groups with prevalences in the 1–2.5\% range, which is large enough to yield precise estimates without substantially increasing recall error. At this sample size, a prevalence of 1\% results in only a 1.09\% chance that no respondents belong to the probe group. At smaller spatial scales, larger prevalences may be necessary, with the mitigating factor that recall error may decline as social networks become more locally concentrated. Additionally, surveys may use restricted definitions of knowing someone (e.g., a smaller time-frame) to reduce recall error when working with larger probe groups. Future work should aim to formalize the trade-off between variability from recall error and the precision of direct probe group estimates.

We suspect it is still valuable to pool direct probe group estimates across regions using a mixed-effect model like the one we employed here. By sharing information, the resulting direct probe group estimates are more robust to the randomness of region-specific samples. Additionally, we were able to easily leverage the uncertainty of the estimates through posterior samples, carrying this uncertainty downstream through the NSUM model. This is similar to the model-based estimator proposed by \cite{feehan2022survey}, and we suggest that a similar method can be used for design-based estimators using a bootstrap approach.

Our primary finding is that direct probe group estimates yield reliable inference. For studies focusing on smaller geographic areas or settings where known population sizes are difficult to obtain, this approach provides a practical alternative. Probe groups are also simpler to select, and even when some known sizes exist, researchers are often limited in their options, making it difficult to choose representative groups. The direct probe group method allows researchers to apply any probe group criteria, including those recommended by \cite{mccormick2010many}. Consistent with \cite{feehan2022survey}, we find it straightforward to work with occupations, as respondents are generally willing to report their occupation, and a representative set spanning age, gender, and wealth categories can be selected.

\bibliographystyle{apalike}
\bibliography{NSUM_bib}

\newpage
\appendix
\appendixpage

{\renewcommand{\arraystretch}{0.92}
\begin{longtable}{|L{0.5\textwidth}|L{0.2\textwidth}|L{0.25\textwidth}|}
\caption{Probe groups, group types, and availability of Governorate-level data used in the Jordan survey. Target groups are included in italics. ``MW-H'' denotes migrant workers working in the hospitality and tourism industry.}
\label{tab:probe_groups} \\
\hline
\textbf{Probe Group} & \textbf{Group Type} & \textbf{Available at Governorate-Level}  \\
\hline
\endfirsthead

\multicolumn{3}{l}{\textit{Table \thetable{} continued from previous page}} \\
\hline
\textbf{Probe Group} & \textbf{Group Type} & \textbf{Available at Governorate-Level}  \\
\hline
\endhead

\hline \multicolumn{3}{r}{\textit{Continued on next page}} \\
\endfoot

\hline
\endlastfoot

Medical doctor & Occupation & No  \\
Lawyer & Occupation & Yes  \\
Basic teacher & Occupation & No  \\
Academic or vocational secondary teacher & Occupation & No  \\
Bank worker & Occupation & No \\
Supermarket employee & Occupation & No  \\
Construction worker & Occupation & No  \\
Engineer & Occupation & Yes  \\
University student & Occupation & No \\
Retired & Occupation & No  \\
Guard & Occupation & No  \\
Other (respondent provided occupation) & Occupation & No \\
Men aged 20-24 & Gender/Age & Yes \\
Women aged 20-24 & Gender/Age & Yes \\
Men aged 65 or older & Gender/Age & Yes \\
Women aged 65 or older & Gender/Age & Yes  \\
People who died in 2023 & Life event & Yes  \\
Grooms who married in 2023 & Life event & Yes  \\
Women who went through a divorce in 2023 & Life event & Yes \\
\textit{MDW}  & Target  &    \\
\textit{Male MDW}  & Target  &    \\
\textit{Female MDW}  &  Target &     \\
\textit{MDW from Philippines}  & Target  &     \\
\textit{MDW from Bangladesh} &  Target &     \\
\textit{MDW from Sri Lanka} &  Target &     \\
\textit{MDW from Indonesia} & Target  &     \\
\textit{MDW from Ethiopia} & Target  &     \\
\textit{MDW from Kenya} & Target  &     \\
\textit{MDW from Ghana} & Target  &     \\
\textit{MDW from Egypt} & Target  &     \\
\textit{MDW from Uganda} &  Target &     \\
\textit{MDW from other countries} & Target  &    \\
\textit{MW-Hosp} &  Target &     \\
\textit{MW-Hosp from Philippines} & Target  &     \\
\textit{MW-Hosp from Bangladesh} &  Target &     \\
\textit{MW-Hosp from Nepal }& Target  &     \\
\textit{MW-Hosp from Uganda} & Target  &     \\
\textit{MW-Hosp from Egypt} &  Target &    
\end{longtable}
}

\newpage

\section{NSUM Questionnaire in Jordan}

Here we provide a redacted (for anonymity) version of the questionnaire used in the Jordan NSUM study.



\section*{Supplementary material}

| | |
|---|---|
| ▇▇▇▇▇▇▇▇▇▇▇▇▇▇▇▇▇▇▇▇▇▇ ▇▇▇▇▇▇▇▇▇▇▇▇▇▇▇▇▇▇ | |
| ▇▇▇▇▇▇▇▇▇▇▇▇▇▇▇▇▇▇▇▇▇▇▇▇▇▇▇▇▇ | |
| ▇▇▇▇▇▇▇▇▇▇▇▇▇▇▇▇▇▇▇▇▇▇▇ | |
| ▇▇▇▇▇▇▇▇▇▇▇▇▇▇▇▇▇▇▇▇▇▇ | |
| ▇▇▇▇▇▇▇▇▇▇▇▇▇▇▇▇▇▇▇▇▇▇ | |
| ▇▇▇▇▇▇▇▇▇▇▇▇▇▇▇▇▇▇▇▇▇▇ ▇▇▇▇▇▇▇▇▇▇▇▇▇▇▇▇▇▇▇▇▇▇ | ▇▇▇▇▇▇▇▇▇▇▇▇▇▇▇▇▇▇▇▇▇▇▇▇▇▇ ▇▇▇▇▇▇▇▇▇▇▇▇▇▇▇▇▇▇▇▇▇▇ ▇▇▇▇▇▇▇▇▇▇▇▇▇▇▇▇▇▇▇▇▇▇▇▇▇▇ ▇▇▇▇▇▇▇▇▇▇▇▇▇▇▇▇▇▇▇▇ ▇▇▇▇▇▇▇ ▇▇▇▇▇▇▇▇▇▇▇▇▇▇▇▇▇▇ |
| ▇▇▇▇▇▇▇▇▇▇▇▇▇▇▇▇▇▇▇▇ ▇▇▇▇▇▇▇▇▇▇▇▇▇▇▇▇▇▇▇▇ | |
| ▇▇▇▇▇▇▇▇▇▇▇▇▇▇▇▇▇▇▇▇▇ ▇▇▇▇▇▇▇▇▇▇▇▇▇▇▇▇▇▇▇▇▇ | |
| ▇▇▇▇▇▇▇▇▇▇▇▇▇▇▇▇▇▇▇▇▇ ▇▇▇▇▇▇▇▇▇▇▇▇▇▇▇▇▇▇▇▇▇ | |
| ▇▇▇▇▇▇▇▇▇▇▇▇▇▇▇▇▇▇▇▇▇ ▇▇▇▇▇▇▇▇▇▇▇▇▇▇▇▇▇▇▇▇▇ | |
| ▇▇▇▇▇▇▇▇▇▇▇▇▇▇▇▇▇▇ | |
| SHELL_COUNTRY. Country | ( ) Afghanistan  ( ) Albania  ( ) Algeria  ( ) Angola  ( ) Argentina<br>( ) Aruba  ( ) Australia  ( ) Austria  ( ) Bahamas  ( ) Bahrain<br>( ) Bangladesh  ( ) Belgium  ( ) Benin  ( ) Bolivia<br>( ) Bosnia and Herzegovina  ( ) Botswana  ( ) Brazil  ( ) Bulgaria<br>( ) Burkina Faso  ( ) Cambodia  ( ) Cameroon  ( ) Canada  ( ) Chile<br>( ) China  ( ) Colombia  ( ) Congo Brazzaville  ( ) Costa Rica<br>( ) Cote d'Ivoire  ( ) Croatia  ( ) Curacao  ( ) Cyprus<br>( ) Czech Republic  ( ) Denmark  ( ) Dominican Republic<br>( ) Ecuador  ( ) Egypt  ( ) El Salvador  ( ) Eritrea  ( ) Estonia<br>( ) Ethiopia  ( ) Finland  ( ) France  ( ) Georgia  ( ) Germany<br>( ) Ghana  ( ) Greece  ( ) Guatemala  ( ) Guinea Bissau<br>( ) Guinea Conakry  ( ) Honduras  ( ) Hong Kong  ( ) Hungary<br>( ) Iceland  ( ) India  ( ) Indonesia  ( ) Iran  ( ) Iraq  ( ) Ireland  ( ) Italy<br>( ) Jamaica  ( ) Jordan  ( ) Kazakhstan  ( ) Kenya  ( ) Kosovo  ( ) KSA<br>( ) Kuwait  ( ) Kyrgyzstan  ( ) Latvia  ( ) Lebanon  ( ) Lesotho<br>( ) Liberia  ( ) Liechtenstein  ( ) Lithuania  ( ) Luxembourg<br>( ) Macedonia  ( ) Malawi  ( ) Malaysia  ( ) Mali  ( ) Malta  ( ) Mexico<br>( ) Montenegro  ( ) Morocco  ( ) Mozambique  ( ) Myanmar<br>( ) Namibia  ( ) Netherlands  ( ) Nicaragua  ( ) Nigeria<br>( ) Northern Ireland  ( ) Norway  ( ) Oman  ( ) Pakistan  ( ) Panama<br>( ) Paraguay  ( ) Peru  ( ) Philippines  ( ) Poland  ( ) Portugal<br>( ) Puerto Rico  ( ) Qatar  ( ) Romania  ( ) Russia  ( ) Rwanda<br>( ) Senegal  ( ) Serbia  ( ) Singapore  ( ) Sint Maarten  ( ) Slovakia<br>( ) Slovenia  ( ) Somalia  ( ) South Africa  ( ) South Korea<br>( ) South Sudan  ( ) Spain  ( ) Sri Lanka  ( ) St Lucia  ( ) Sudan<br>( ) Suriname  ( ) Swaziland  ( ) Sweden  ( ) Switzerland  ( ) Taiwan<br>( ) Tanzania  ( ) Thailand  ( ) Tunisia  ( ) Türkiye  ( ) UAE  ( ) Uganda<br>( ) Ukraine  ( ) United Kingdom  ( ) United States  ( ) Uruguay<br>( ) Uzbekistan  ( ) Venezuela  ( ) Vietnam  ( ) Yemen  ( ) Zambia<br>( ) Zimbabwe |
| SHELL_LANGUAGE. Language selection | ( ) English  ( ) Assamese  ( ) Bengali  ( ) Gujarati  ( ) Hindi<br>( ) Kannada  ( ) Malayalam  ( ) Marathi  ( ) Oriya  ( ) Tamil  ( ) Telugu<br>( ) Spanish  ( ) Polish  ( ) Arabic  ( ) French  ( ) Igbo  ( ) Hausa<br>( ) Yoruba  ( ) Swahili  ( ) Acholi  ( ) Ateso  ( ) Luganda  ( ) Luo<br>( ) Runyankole/Rukiga  ( ) Runyoro/Rutoro  ( ) Portuguese<br>( ) Chinese  ( ) Tagalog  ( ) Cebuano  ( ) Ilonggo  ( ) Ilokano<br>( ) Bicolano  ( ) Waray  ( ) Thai  ( ) Vietnamese  ( ) Baoulé  ( ) Bété<br>( ) Dioula  ( ) Senoufo  ( ) Akan  ( ) Dagbani  ( ) Ga  ( ) Ewe  ( ) Farsi<br>( ) Russian  ( ) Urdu  ( ) Indonesian  ( ) Malay  ( ) Macua  ( ) Ndau |

| | |
|---|---|
| | ( ) Sena  ( ) Tsonga  ( ) Bemba  ( ) Kaonde  ( ) Lozi  ( ) Lunda<br>( ) Luvale  ( ) Nyanja  ( ) Tonga (Zambezi)  ( ) Hong Kong Cantonese<br>( ) Korean  ( ) Taiwanese Mandarin  ( ) Kurdish  ( ) Bulgarian<br>( ) Romanian  ( ) Kazakh  ( ) Ukrainian  ( ) Croatian  ( ) Serbian<br>( ) Albanian  ( ) Hungarian  ( ) Dutch  ( ) German  ( ) Catalan<br>( ) Spanish - Spain  ( ) Amharic  ( ) Wolof  ( ) Bosnian<br>( ) Macedonian  ( ) Montenegrin  ( ) Slovenian  ( ) Tigrinya  ( ) Greek<br>( ) Turkish  ( ) Czech  ( ) Slovak  ( ) Danish  ( ) Estonian  ( ) Finnish<br>( ) Swedish  ( ) Italian  ( ) Latvian  ( ) Lithuanian  ( ) Luxembourgish<br>( ) Maltese  ( ) Slovak  ( ) Norwegian  ( ) German - Austria  ( ) Oromo<br>( ) Karamojong  ( ) Runyakitara  ( ) Lugbara  ( ) Somali  ( ) Afrikaans<br>( ) Zulu  ( ) Xhosa  ( ) Sotho  ( ) Sepedi  ( ) Lingala  ( ) Kinyarwanda<br>( ) Arabic Juba  ( ) Siswati  ( ) Setswana  ( ) Burmese  ( ) Sinhalese<br>( ) Georgian  ( ) Dari (Afghanistan)  ( ) Pashto (Afghanistan)<br>( ) Kyrgyz  ( ) Uzbek  ( ) Khmer  ( ) English - United Kingdom<br>( ) Tajik  ( ) Spanish - Peru |
| SHELL_LANGUAGECODE. Language Code | |
| SHELL_INTRO_GDPR. Good morning/afternoon/evening, I am _____ from ▉▉▉▉▉▉▉▉▉▉▉▉<br>We are currently conducting a research study and your responses could help to shape and to estimate migrant workers presence in Jordan. There are no right or wrong answers, and please be assured that the information collected from you will be treated confidentially. Answers from all participants will be combined and only anonymized results will be used for research reporting.<br><br>Are you happy to proceed with the interview? | ( ) Yes  ( ) No |
| SHELL_RECORDING_CONFIRMATION. For quality purposes some of the questions during the interview will be automatically audio recorded.<br><br>Do you agree with the audio recording? | ( ) Yes  ( ) No |
| SHELL_PRIVACY_NOTICE. INTERVIEWER INSTRUCTION: Give the privacy notice information to the respondent and read the following sentence.<br><br>I'm bringing to your attention the information from this privacy notice which explains about the legal basis and the purposes for processing your personal data as well as your rights under data protection regulations. | ( ) Yes  ( ) No |
| SHELL_SIGNATURE. In that case could you please sign here to confirm your consent to proceed with participating in the interview.<br><br>By signing electronically below, I acknowledge that I have consented to proceed with participating in the interview | |
| Q.Supervisor. Who is your supervisor | |

| | | |
|---|---|---|
| | | |
| SHELL_GENDER. Gender | ( ) Male  ( ) Female | |
| *BY OBSERVATION* | | |
| SHELL_AGE. What was your age on your last birthday? | ( ) Exact Age  ( ) Refused | |
| SHELL_AGE_RECODED. Age groups | ( ) 10-15 years old  ( ) 16-24 years old  ( ) 25-34 years old<br>( ) 35-44 years old  ( ) 45-54 years old  ( ) 55-64 years old<br>( ) 65+ years old | |
| SHELL_INFO_PROJECTSCRIPT_START. Scripter, add the project specific questions after this info item and before the SHELL_CLOSE question. | | |
| 2a. What governorate do you reside in? | ( ) Irbid  ( ) Ajloun  ( ) Jerash  ( ) Mafraq  ( ) Balqa  ( ) Amman<br>( ) Zarqa  ( ) Madaba  ( ) Karak  ( ) Tafilah  ( ) Ma`an  ( ) Aqaba | |
| *(select one option)* | | |
| 3. What is your nationality? | ( ) Jordanian  ( ) Other specify | |
| 3a. Which of these occupations do you hold | [ ] Medical doctor  [ ] Lawyer  [ ] Basic teacher<br>[ ] Academic or Vocational Secondary teacher  [ ] Bank worker<br>[ ] Supermarket employee  [ ] Construction worker  [ ] Engineer<br>[ ] University student  [ ] Retired  [ ] Guard  [ ] Other | |
| **Probe Groups**<br>We would like to ask you some questions about people that you know. People you know should satisfy all of the three following criteria:<br>• You know them by sight and name, and they also know you by sight and name.<br>• You have had some contact with them – either in person, over the phone, on WhatsApp, or on the computer – in the past 12 months<br>• They live in the same Governorate you currently reside in<br>These people could be your relatives, friends, colleagues, etc . | | |
| 4. How many men do you know who are aged 20-24? | | |
| 5. How many women do you know who are aged 20-24? | | |
| 6. How many men do you know who are aged 65 or older? | | |
| 7. How many women do you know who are aged 65 or older? | | |
| 8. How many people do you know who died in 2023 | | |
| 9. How many grooms do you know who married in 2023? | | |

| | |
|---|---|
| 9.a)Did you marry in 2023? | ( ) Yes  ( ) No |
| 10.How many women do you know who went through a divorce in 2023? | |
| b)Did you go through a divorce in 2023? | ( ) Yes  ( ) No |
| 11.How many people do you know who are medical doctors? | |
| 12.How many people do you know who are lawyers? | |
| 13.How many people do you know who are basic teachers? | |
| 14.How many people do you know who are academic or vocational secondary teachers? | |
| 15.How many people do you know who are bank workers? | |
| 16.How many people do you know who are supermarket employees? | |
| 17.How many people do you know who are construction workers? | |
| 18.How many people do you know who are engineers? | |
| 19.How many people do you know who are university students? | |
| 20.How many people do you know who are retired? | |
| We would like to ask you some additional questions about people that you know who are Migrant Domestic Workers. Migrant domestic Workers refer to any non-Jordanian person, male or female, engaged in domestic work within an employment relationship (e.g.: working within a household and performing a variety of household services such as cleaning, household maintenance, cooking, laundry and ironing, household errands, care for children or elderly dependents, driving or guarding the house. | |
| 21.How many people do you know who are migrant domestic workers? | |
| i would like to ask a couple of additional questions about these migrant domestic workers you know. | |
| 22.How many of those migrant domestic workers are male? | |
| 23.How many of those migrant domestic workers are female? | |
| 24.How many of those migrant domestic workers are from the Philippines? | |
| 25.How many of those migrant domestic workers are from Bangladesh? | |
| 26.How many of those migrant domestic workers are from Sri Lanka? | |
| 27.How many of those migrant domestic workers are from Indonesia? | |
| 28.How many of those migrant domestic workers are from Ethiopia? | |

| | |
|---|---|
| 29. How many of those migrant domestic workers are from Kenya? | |
| 30. How many of those migrant domestic workers are from Ghana? | |
| 31. How many of those migrant domestic workers are from Egypt? | |
| 32. How many of those migrant domestic workers are from Uganda? | |
| 33. How many of those migrant domestic workers are from any country other than those asked about above? | |
| Q33a what are these nationalities | [ ] 1 [ ] 2 [ ] 3 [ ] 4 [ ] 5 [ ] 6 [ ] 7 [ ] 8 [ ] 9 [ ] 10 |
| Q33b How many of those migrant domestic workers that you know are from? | |
| I would now like to ask you questions about the migrant workers you know in the hospitality and tourism sector | |
| Migrant workers working in the hospitality or tourism sector refer to any non-Jordanian person, male or female, engaged in the hotels and tourism facilities in Jordan (hotels, airlines, travel agencies, touristic restaurants and buffets, resorts, bars and nightclub...) | |
| 34. How many people do you know who are migrant workers in the hospitality sector? | |
| 35. How many of those migrant workers in the hospitality sector are from the Philippines? | |
| 36. How many of those migrant workers in the hospitality sector are from Bangladesh? | |
| 37. How many of those migrant workers in the hospitality sector are from Nepal? | |
| 38. How many of those migrant workers in the hospitality sector are from Uganda? | |
| 39. How many of those migrant workers in the hospitality sector are from Egypt? | |
| 40. Now we have reached the end of the interview, do you have any comments you would like to share with us? | |
| Term. Are you sure you want to terminate? | ( ) Yes  ( ) No |
| SHELL_CLOSE. Thank you very much for participating in our survey. We sometimes re-contact respondents to verify that the survey was indeed carried out. For this purpose we require some contact details from you in order to be able to check this should your interview be selected for quality control. | |
| SHELL_DDG_STATUS. SHELL_DDG_STATUS | |
| *DDG only question to validate survey* | |
| SHELL_NAME. Respondent name: | |
| SHELL_BLOCK_TEL. Telephone number(s) | |
| SHELL_MOBTEL. Mobile telephone number: | |



\end{document}